\documentclass[11pt]{article}                               
\pdfoutput=1
\usepackage{latexsym}
\usepackage{amssymb}
\usepackage{mathrsfs}
\usepackage{amsmath}
\usepackage{amsthm}
\usepackage{color}
\usepackage{graphicx}

\def\AFOUR{%
\setlength{\textheight}{9.0in}%
\setlength{\textwidth}{5.75in}%
\setlength{\topmargin}{-0.375in}%
\hoffset=-.5in%
\renewcommand{\baselinestretch}{1.17}%
\setlength{\parskip}{6pt plus 2pt}%
}

\AFOUR                                           

\makeatletter
\def\section{\@startsection {section}{1}{\z@}{-3.5ex plus -1ex minus
 -.2ex}{2.3ex plus .2ex}{\large\bf}}
\def\subsection{\@startsection{subsection}{2}{\z@}{-3.25ex plus -1ex minus
 -.2ex}{1.5ex plus .2ex}{\normalsize\bf}}
\makeatother

\makeatletter
\@addtoreset{equation}{section}

\makeatother

\newcommand{\nc}{\newcommand}
\newcommand{\rnc}{\renewcommand}

\nc{\bea}{\begin{eqnarray}}
\nc{\eea}{\end{eqnarray}}
\nc{\be}{\bea}
\nc{\ee}{\eea}

\rnc{\a}{\alpha}
\nc{\ab}{\bar{\a}}
\nc{\ap}{\a^{+}}
\nc{\abm}{\ab^{-}}
\rnc{\b}{\beta}
\nc{\bb}{\bar{\b}}
\nc{\bbp}{\bb_{\zb}^{+}}
\nc{\bm}{\b_{z}^{-}}
\nc{\oa}{\overline{\a}}
\nc{\ob}{\overline{\b}}
\rnc{\gg}{\gamma}
\rnc{\d}{\delta}
\nc{\f}{\phi}
\nc{\fb}{\bar{\phi}}
\nc{\vf}{\varphi}
\nc{\p}{\psi}

\rnc{\c}{\chi}
\nc{\la}{\lambda}
\nc{\m}{\mu}
\nc{\n}{\nu}
\rnc{\o}{\omega}
\nc{\Om}{\Omega}
\rnc{\t}{\theta}
\nc{\eps}{\epsilon}
\rnc{\S}{\Sigma}
\nc{\F}{\Phi}

\nc{\trac}[2]{{\textstyle\frac{#1}{#2}}}

\nc{\ex}[1]{\mbox{e}^{\,\textstyle#1}}

\nc{\mat}[4]{\left(\begin{array}{cc}#1&#2\\#3&#4\end{array}\right)}

\nc{\som}[9]{\left(\begin{array}{ccc}#1&#2&#3\\#4&#5&#6\\#7&#8&#9%
\end{array}\right)}

\nc{\tr}{\mathop{\mbox{tr}}\nolimits}
\nc{\ad}{\mathop{\mbox{ad}}\nolimits}
\nc{\Tr}{\mathop{\mbox{Tr}}\nolimits}
\nc{\Det}{\mathop{\mbox{Det}}\nolimits}
\nc{\rk}{\mathop{\mbox{rk}}\nolimits}
\nc{\ra}{\rightarrow}
\nc{\Ra}{\Rightarrow}
\nc{\LRa}{\Leftrightarrow}
\nc{\ot}{\otimes}
\nc{\nul}{\noindent\underline}
\nc{\non}{\nonumber\\}

\nc{\subs}[1]{{\vspace*{0.5cm}}%
{\noindent\underline{#1}}{\addcontentsline{toc}{subsection}{#1}}%
{\vspace*{0.3cm}}}

\nc{\zb}{\bar{z}}
\rnc{\lg}{\mathfrak{g}}
\nc{\lt}{\mathfrak{t}}
\nc{\lk}{\mathfrak{k}}
\nc{\lh}{\mathfrak{h}}
\nc{\pik}{\Pi_{\lk}}
\nc{\pip}{\Pi_{+}}
\nc{\pim}{\Pi_{-}}
\nc{\pih}{\Pi_{\lh}}
\nc{\jz}{J_{z}}
\nc{\jzh}{\jz^{\lh}}
\nc{\jzp}{\jz^{+}}
\nc{\jzm}{\jz^{-}}
\nc{\del}{\partial}
\nc{\dz}{\del_{z}}
\nc{\dzb}{\del_{\bar{z}}}
\nc{\az}{A_{z}}
\nc{\azb}{A_{\bar{z}}}
\nc{\g}{g^{-1}}
\nc{\dw}{\Delta_{W}}
\nc{\Ad}{{\mbox{Ad}}}
\nc{\ks}{Ka\-za\-ma-\-Su\-zu\-ki}
\nc{\KS}{\ks}
\nc{\ksm}{\ks\ model}
\rnc{\AA}{{\Bbb A}}
\nc{\BB}{{\Bbb B}}
\nc{\CC}{{\mathbb{C}}}
\nc{\PP}{{\Bbb P}}
\nc{\cpm}{\CC\PP(m)}
\nc{\cpn}{\CC\PP(n)}
\nc{\cp}[1]{\CC\PP(#1)}
\nc{\gmn}{G(m,m+n)}
\nc{\gmnk}{\gmn_{k}}
\nc{\cO}{{\cal O}}
\nc{\bcO}{\bar{\cO}}
\nc{\bO}{\bar{O}}
\nc{\oQ}{\overline{Q}}

\begin{document}
\global\parskip=4pt

\makeatletter
\begin{titlepage}
\begin{center}
{\LARGE\bf Chern-Simons Theory with Complex Gauge\\[.2in]
  Group on Seifert Fibred 3-Manifolds}\\

\vskip 0.7in
{\bf Matthias Blau}
\vskip .1in
Albert Einstein Center for Fundamental Physics\\
Institute for Theoretical Physics\\ University of Bern, Switzerland.
\vskip 0.2in
\vskip 0.2in
{\bf George Thompson}
\vskip .1in
Abdus Salam ICTP \\
 Trieste, 
Italy.\\

\end{center}
\vskip .4in
\begin{abstract} 
\noindent 
We consider Chern-Simons theory with complex gauge group and present a
complete non-perturbative evaluation of the path integral (the partition
function and certain expectation values of Wilson loops) on Seifert
fibred 3-Manifolds. We use the method of Abelianisation. In certain cases
the path integral can be seen to factorize neatly into holomorphic and
anti-holomorphic parts. We obtain closed formulae of this factorization
for the expectation values of torus knots.

\end{abstract}

\end{titlepage}
\makeatother


\setcounter{footnote}{0}

\section{Introduction}

Chern-Simons theory with a complex gauge group $G_{\mathbb{C}}$
(viewed as a complexification of some compact gauge group $G$)
 has been under study since it was first considered in
 \cite{Witten-Complex-CS}. Formal aspects of gauge fixing for
 semi-simple gauge groups appeared in \cite{Bar-Natan-Witten} and a
 Hamiltonian version of quantization appears in \cite{BNP}. In the
 case of $G_{\mathbb{C}}=SL(2, \mathbb{C})$ Chern-Simons theory
 appears as a gravitational theory in three dimensions \cite{WBF,
   Gukov} in first order formalism. The `volume conjecture' which
relates certain (limits of) quantum knot invariants to the hyperbolic
volume of the knot complement \cite{Kashaev, MMOTY} renewed interest
in complex Chern-Simons theory. This is mainly for hyperbolic
manifolds (or knots). 

Rather more recently the 3d-3d correspondence \cite{DGH} and \cite{DGG1,
DGG2} has revived the study of Chern-Simons theory with complex gauge
group and in particular requires one to know the partition function
on manifolds which are not necessarily hyperbolic. Short of having an
ab initio derivation, many of the results are in fact conjectural and
are used to test, rather than to prove the veracity of, the proposed
correspondence.

In the study of the quantum theory canonical quantization of the
complexified theory has its 
difficulties. The phase space is the symplectic manifold of flat
$G_{\mathbb{C}}$ connections on a Riemann surface $\Sigma$ up to gauge
transformations. That
moduli space has the standard description
\be
\mathcal{M}_{\mathbb{C}}(\Sigma) = \mathrm{Hom}(\pi_{1}(\Sigma), \,
G_{\mathbb{C}})/\Ad G_{\mathbb{C}}\label{complex-flat}
\ee
As explained by Witten \cite{Witten-Complex-CS},
rather than quantizing
this non-compact space,one alternative is to consider instead the space of flat $G$
connections, up to gauge equivalence,
\be
\mathcal{M}(\Sigma) = \mathrm{Hom}(\pi_{1}(\Sigma), \,
G)/\Ad G
\ee
and square integrable sections of a pre-quantum line bundle over this
space. Dealing directly with the quantization of (\ref{complex-flat})
has been pushed forward in \cite{Dimofte-Gukov} and in particular in
\cite{Gukov-Pei}, where an integral grading of the infinite
dimensional Hilbert space into finite dimensional subspaces
has been given. 

One of our aims here is to instead perform the path integral
directly (foregoing the canonical procedure completely). In a series
of papers we have evaluated the Chern-Simons partition 
function with compact and simply connected gauge group on (and
invariants for certain links in) Seifert fibered
3-manifolds \cite{BT-CS, BT-S1, BT-Seifert} (by a suitable gauge
fixing and pushing the problem 
down to a 2-dimensional Abelian gauge theory).  While we have applied
these ideas to three dimensional $BF$ theory \cite{WBF,tftpr} in
\cite{BT-S1,BTO} we have not dealt with the complex case.

The computations that we perform are for Seifert
3-manifolds and Seifert
manifolds are never hyperbolic. There are, however, advantages to this
choice of 3-manifold. For the most part the situation here is vastly
simplified as (with suitable gauge fixing) almost all of 
the path integrals that we encounter are Gaussian (with sources). The
final answer for the partition function with complex gauge group 
takes a very simple form (\ref{finite-partition3}) as 
an integral over the complex Cartan subalgebra and a summation whose
range depends on the order $|d|$ of $\mathrm{H}_{1}(M, \mathbb{Z})$,
\be
Z_{M}[t, \, \overline{t}] =\frac{1}{|W|}
\sum_{1
  \leq\mathbf{n}\leq \mathbf{d}}
\int_{\lt_{\mathbb{C}}} d^{2\mathbf{rk}(G)}z\, 
\exp{\left(iI^{\mathbf{n}}(t, \, \overline{t}) \right)}\;\;
\widehat{\tau}_{M}^{1/2}(z;\, a_{1}, \dots , a_{N}) \label{intro-form}
\ee
The integrand involves the Ray-Singer torsion $\widehat{\tau}_{M}$ of
the 3-manifold with connections in the complexified group and the
exponent $I^{\mathbf{n}}(t, \, \overline{t})$ is a polynomial of
degree two in the integration variables. The complex parameter
$t=k+is$ is the complex `level' or coupling constant of the
theory. The circle $V$-bundle of the line $V$-bundle $\mathcal{L}$ over
$S^{2}$ with $N$ orbifold points is the
3-manifold $M$ in question. Note that (\ref{intro-form}) is very
similar to (a square of) the general formula given \cite{Marino},
except that we have an integral rather than a sum to
perform over one of the factors. The continuous variable reflects the
non-compact nature of the gauge group.

Another motivation for this work is to understand the holomorphic
factorization of $G_{\mathbb{C}}$ Chern-Simons theory
\cite{Witten-Bonn}. The path integral under consideration may be
written as
\be
Z[M, \, t, \, \overline{t}] = \int D \mathscr{C}D\overline{\mathscr{C}}
\,\, \ex{\left(itI(\mathscr{C}) +i\overline{t}I(\overline{\mathscr{C}}) \right)}
\ee
where the overline indicates complex conjugation. The natural question
is how close are we to being able to make sense of a factorization of
the path integral into holomorphic and anti-holomorphic parts
\be
Z[M, \, t, \, \overline{t}] \stackrel{?}{=} Z[M, \, t]\, . \, \overline{Z[M,\,
t] }\label{fact}
\ee
where
\be
Z[M, \, t]  =  \int D \mathscr{C}\,\, \ex{\left(itI(\mathscr{C})
    \right)} \;\;\;\; \mathrm{and} \;\; \;\;\;
\overline{Z[M,\,t] } =  \int D \overline{\mathscr{C}}\,\, \ex{
  \left( i\overline{t}I(\overline{\mathscr{C}})
    \right)} \nonumber
\ee
and if such partition functions are to make sense what `contour'
path integrals are being performed on the right hand side? It turns out
that around a given flat connection (\ref{fact}) holds in perturbation
theory. In \cite{DGLZ}, equation (1.6), and \cite{Witten-Bonn},
equation (2.27), it is shown that the
correct formula is 
\be
Z[M, \, t, \, \overline{t}] = \frac{1}{|W|}\sum_{\rho, \overline{\rho}}
n_{(\rho, \, \overline{\rho})}\,\, Z^{\rho}[M, \,
t]\, . \, \overline{Z^{\rho}[M,\, t]}\label{Witten-factor}
\ee
for some numbers $n_{(\rho, \, \overline{\rho})}$, where $\rho$
labels the solution to the flatness equation, and for particular
contours. We will discuss below, under which circumstances 
and in which sense our exact non-perturbative result 
\eqref{intro-form} displays this holomorphic factorization.

In the next section we introduce Chern-Simons theory with a complex
gauge group proper, taking it to be the complexification
$G_{\mathbb{C}}$ of a compact gauge group $G$. 
There we point out that there exists a symmetry of
the path integral under the exchange $t
\leftrightarrow \widehat{t}$ which in part motivates holomorphic
factorization. Various formal limits of the coupling constants $k$ and $s$
are considered in section \ref{various-limits}. These limits allow us
to get an understanding of what may be gleaned from allowing for a
complex gauge group. Sections \ref{section-gf} and \ref{section-seifert} are
devoted to our choice of gauge and choice of manifold. We note that
the choice of gauge is unitary and so avoids the pitfalls outlined in
\cite{Bar-Natan-Witten} of the more usual Lorentz gauge. 

In section \ref{section-2d} we integrate out the fibre dependence of
all the fields leaving us with an effective two dimensional Abelian
gauge theory (\ref{ab-partition}). This theory can also be seen to
formally factorize into holomorphic and anti-holomorphic
components and we see that the Ray-Singer torsion enjoys this
factorization. In section \ref{section-finite} we integrate all
non-constant modes in the two dimensional theory to be left with the
finite dimensional integral (\ref{intro-form}). This integral is such that the
integrand factorises into holomorphic and anti-holomorphic parts
itself and so we have been able to follow the factorization at each step of the
evaluation of the path 
integral. In particular for (\ref{intro-form}) factorization, at least
for certain $M$, is based
upon the fact that the following integral 
\be
\int d^{2}z \, \exp{\left(az^{2} + b\overline{z}^{2} + cz
    +d\overline{z} \right)} = \frac{\pi}{\sqrt{-ab}}\, \exp{\left(-
    \frac{c^{2}}{4a} - \frac{d^{2}}{4b} \right)}\label{Gaussian}
\ee
may be expressed as the product of two contour
integrals (not closed contours)
\be
\ex{\left( -i\pi/2\right)}\, .\, \int_{\Gamma}dz \, \exp{\left(az^{2}+cz\right)}\, .\,
\int_{\Gamma'}d\overline{z} 
\, \exp{\left(b\overline{z}^{2}+ d\overline{z}\right)} \label{contours}
\ee
for example with both $\Gamma$ and $\Gamma'$ being the real
axis. These last formulae immediately imply that the `holomorphic'
factor of the $G_{\mathbb{C}}$ theory is just the partition function
of the $G$ theory (at least up to framing and a change of
level) once we sum over $\mathbf{n}$. 

Given that we have established factorization the question of what we
are summing over (or what does $\mathbf{n}$ represent geometrically) arises. In
section \ref{section-flat}, by comparing 
with expectations of the $s\rightarrow \infty$ limit, we are also able
to show that, at least in that limit, the sum is over particular flat
Abelian connections. At least for the Lens spaces this is in keeping
with the expectations of \cite{DGLZ, Dimofte-Gukov, Witten-Bonn}.

While one can perform the integral (\ref{intro-form}) exactly (we have
done it for the Lens spaces) it is instructive to use the
factorization formula (\ref{contours}) to evaluate the holomorphic and
anti-holomorphic parts separately. As expected, up to framing dependent
phases and a shift in the level,
these calculations agree for $G_{\mathbb{C}}=SL(2, \mathbb{C})$ with
those of the partition function $SU(2)$. We perform the calculations
on $M(n,0, (a_{1}, b_{1}), (a_{2}, b_{2}))$ (Seifert manifolds with
$N=2$ orbifold points on the base $S^{2}$) giving a redundant
(diffeomorphic) description of large numbers of Lens spaces in section
\ref{section-calcs}. This description allows us to evaluate the
expectation values of Torus knots (in $S^{3}$, though it is also just
as easy to evaluate them in Lens spaces).

Somewhat less obvious is factorization for Seifert manifolds with
$N\geq 3$ exceptional fibres. We do not claim to have shown this,
though (\ref{intro-form}) holds. However, in the $s\rightarrow \infty$
limit holomorphic factorization holds (with the holomorphic theory
being a kind of `square root' of $BF$ theory).

For the future, we note that the techniques used here can equally well
be used to evaluate the partition functions and observables in
Yang-Mills theory with complex gauge group and the
$G_{\mathbb{C}}/G_{\mathbb{C}}$ model (at least on $S^{2}$) and for
supersymmetric theories.

\section{Complex Chern-Simons Theory}
Let $G$ denote a simple and simply connected compact Lie group and let
$G_{\mathbb{C}}$ be its complexification. Consider a 3-manifold $M$
and the trivial $G_{\mathbb{C}}$ bundle over it $P_{\mathbb{C}}= M \times
G_{\mathbb{C}}$. Let $\mathcal{A}_{\mathbb{C}}$ be the affine space of
connections on $P_{\mathbb{C}}$. The Chern-Simons action is,
\bea
I(t, \, \widehat{t}) & = &  \frac{t}{8\pi} \int_{M} \Tr{\left( \mathscr{C}\wedge d \mathscr{C} +
  \frac{2}{3} \mathscr{C}\wedge\mathscr{C}\wedge\mathscr{C}\right)} \nonumber\\
& & \;\;\; + \frac{\widehat{t}}{8\pi} \int_{M}
\Tr{\left(\overline{\mathscr{C}} \wedge d 
  \overline{\mathscr{C}} +   \frac{2}{3}
  \overline{\mathscr{C}} \wedge \overline{\mathscr{C}} \wedge
  \overline{ \mathscr{C}}\right) },\label{ITT}
\eea 
for $\mathscr{C} \in \mathcal{A}_{\mathbb{C}}$ and with $t,
\widehat{t} \in \mathbb{C}$ (but not necessarily complex
conjugates). The over-line indicates minus Hermitian conjugation (which
is the same as complex conjugation within the trace). We may write 
\be
\mathscr{C}= \mathscr{A} + i\mathscr{B}, \;\;\; \mathscr{A}, \,
\mathscr{B} \in \Omega^{1}(M, 
\mathrm{Lie}\, G) \label{CAB}
\ee
where both $\mathscr{A}$ and $\mathscr{B}$ are anti-Hermitian. Under
this split the action becomes
\bea
I(k,s) & = &  \frac{k}{4\pi} \int_{M}  \Tr{\left( \mathscr{A}\wedge d
    \mathscr{A} + \frac{2}{3} \mathscr{A} 
    \wedge \mathscr{A} \wedge \mathscr{A} - \mathscr{B} \wedge
    d_{\mathscr{A}}\mathscr{B} \right)} \nonumber \\ 
&  &  \;\;\;- \frac{s}{2\pi} \int_{M}  \Tr{\left( \mathscr{B} \wedge
    F_{\mathscr{A}} - 
    \frac{1}{3} \mathscr{B} \wedge \mathscr{B} \wedge
    \mathscr{B}\right)} 
\eea
where $t=k+is$, $\widehat{t}=k- is$. We let $I$ denote the Chern-Simons
action for the compact structure group $G$,
\be
I = \frac{1}{4\pi} \int_{M}  \Tr{\left( \mathscr{A}\wedge d
    \mathscr{A} + \frac{2}{3} \mathscr{A} 
    \wedge \mathscr{A} \wedge \mathscr{A}\right)}
\ee
As the exponential of the action
should be invariant under $\mathcal{G} \subset
\mathcal{G}_{\mathbb{C}}$ transformations we must have that $k \in
\mathbb{Z}$, however, this imposes no constraint on $s\in \mathbb{C}$.
Unitarity of the
theory is another issue. As explained in \cite{Witten-Complex-CS}
unitarity for a Euclidean theory amounts to requiring that the
argument of the path integral under a change orientation is the same
as complex conjugation. Consequently, on putting the manifold
dependence in $I(t, \, \widehat{t})$ (\ref{ITT}), 
\be
-\overline{I}(M,t,\widehat{t})= I(-M,t,\widehat{t})
\ee
There are two choices for $s$ which lead to a
unitary theory \cite{Witten-Complex-CS}:
\begin{enumerate}
\item $s \in \mathbb{R}$ with $\mathscr{C} \longrightarrow \mathscr{C}
   $ under orientation reversal or
\item $s \in i\mathbb{R}$ and under a reversal of orientation
  $\mathscr{C} \longrightarrow \overline{\mathscr{C}}$
\end{enumerate}
The unitary
theory that we consider here is the one where $s \in \mathbb{R}$ so
that $\widehat{t} =\overline{t}$.

\subsection{A Symmetry $t  \leftrightarrow \widehat{t}$}
We note that by simply sending $\mathcal{B}\rightarrow -\mathcal{B} $
in (\ref{CAB}) we are exchanging $\mathcal{C} \leftrightarrow
\overline{\mathcal{C}} $ so that the theory (providing one's
regularization preserves this transformation) is invariant under an
interchange of $t  \leftrightarrow \widehat{t}$. This transformation
is independent of the unitarity conditions just described. The
interchange $t  \leftrightarrow \widehat{t}$ amounts to saying that
the theory is invariant under $s \rightarrow -s$.

\section{Various Limits}\label{various-limits}

One may get a feel for what the invariants are by taking various
limits of the parameter $s$ that appears in the theory. This provides 
some intuition
for the formal structure of the theory. For the rest
of this section the 3-manifold $M$ is taken to be a rational homology
sphere ($\mathbb{Q}$HS), $\mathrm{H}_{1}(M , \, \mathbb{Q})=0$, so
that the moduli space 
of flat $G_{\mathbb{C}}$ connections is a set of isolated
points. $M$ being a $\mathbb{Q}$HS means that we do not have
to worry about zero modes associated with flat directions in the path
integral. 
\subsection{The $s \longrightarrow 0$ Limit}
A potential source of zero modes are solutions to the equations 
\be
d_{\mathscr{A}}\, \mathscr{B}=0
\ee
however, by our choice of $M$, we do not have to confront solutions to
this equation about a flat connection $\mathscr{A}$. This means that
$I(k,\,0)$ is a perfectly good action in this case. Performing the
Gaussian Integration over $\mathscr{B}$ (including gauge fixing) gives,
up to a phase,
\be
\sqrt{\tau_{M}(\mathscr{A})}\label{RST-root}
\ee
where $\tau_{M}$ is the Ray-Singer torsion on $M$ of the given flat
connection. The phase that one gets from this path integral
exactly compensates that that one would get on a perturbative
evaluation of the path integral over $\mathscr{A}$
\cite{Bar-Natan-Witten}. 

While it is pleasing to see the Ray-Singer Torsion (\ref{RST-root})
arise, its presence in a non-perturbative treatment of the path integral
complicates matters. The argument of the torsion in general is not a
flat connection and the integral over the space of $G$ connections
$\mathcal{A}$ now has its measure given by (\ref{RST-root}) which is
in principle very complicated.

\subsection{The $s \longrightarrow \infty$ Limit}

To ensure that $\exp{(iI(k,\, s))}$ does not oscillate wildly as $s
\longrightarrow \infty$  we also scale
\be
\mathscr{B} \longrightarrow \mathscr{B}/s\nonumber
\ee
Then the leading terms in the action become
\be
\frac{k}{4\pi} \int_{M}  \Tr{\left( \mathscr{A}\wedge d
    \mathscr{A} + \frac{2}{3} \mathscr{A} 
    \wedge \mathscr{A} \wedge \mathscr{A}\right)} - \frac{1}{2\pi}
\int_{M}  \Tr{\left( \mathscr{B} \wedge 
    F_{\mathscr{A}} \right)} 
\ee
which is a combination of $BF$ and Chern-Simons theories. Had we kept
higher order terms in $1/s$ we would have been able to develop a
perturbation theory in this limit. Formally, the measure of the theory in this
limit together with the scaling (up to $\pi$ factors) goes as
\be
\prod_{i}\left(d \mathscr{C}_{i}\, .\, \sqrt{t}\right) \left(d\overline{
    \mathscr{C}}_{i}\, .\, \sqrt{\widehat{t}}\right) = \prod_{i}\left( d
  \mathscr{A}_{i}d
    \mathscr{B}_{i} \, .\, \sqrt{k^{2}+s^{2}}\right) \longrightarrow
  \prod_{i}\left( d 
  \mathscr{A}_{i}d
    \mathscr{B}_{i} \right) \label{BF-limit}
\ee
for $i$ some complete basis indexing set of the forms that appear in
the path integral. This is not quite right, as we have not taken the
volume of the gauge group into account. A correct analysis will
require us to multiply the measure by a factor so that the
right hand limit in (\ref{BF-limit}) is what one obtains. One can see
the problem directly in (\ref{intro-form}) where the proposed limit
always gives zero. We will come
back to this in Section \ref{section-flat}.

Notice that for consistency in this
limit we need also to make the Wigner-In\"{o}n\"{u} group contraction
\be
G_{\mathbb{C}} \rightarrow IG
\ee
The path integral is straightforward to perform and, at least
formally, one obtains that the partition function is
\be
Z(k, \infty) = \sum_{\mathscr{A} \in \mathcal{M}} \, \exp{\left(ik
    I(\mathscr{A}) \right)}\, . \,
\tau_{M}(\mathscr{A} )\label{large-s}
\ee
where $\mathcal{M}$ is the moduli space of flat $G$ connections on
$M$. One can do better and also formally evaluate the expectation value
of Wilson loops of the $G$ connection $\mathscr{A}$,
\be
\langle \prod_{i} \Tr_{R_{i}}{\left( P. \ex{\int_{\gamma_{i}}
        \mathscr{A}} \right) }\rangle  =  \sum_{\mathscr{A}
  \in \mathcal{M}} \, \exp{\left(ik 
    I(\mathscr{A}) \right)}\, . \, 
 \prod_{i}\Tr_{R_{i}}{\left( P. \ex{\int_{\gamma_{i} }
        \mathscr{A} }\right) }\,\tau_{M}(\mathscr{A} )
\ee
In section 9 of \cite{BT-S1} and in \cite{BTO} we were able to
evaluate the path integral explicitly 
for certain Seifert manifolds (for example the computation gives the
explicit form of the Ray-Singer torsion for each flat connection). One
may also obtain explicit expressions with the insertion of
Wilson loops which wrap along the fibre of $M$ and the class of Wilson loops
include those involving $\mathscr{B}$ with finite and infinite dimensional
representations \cite{BTO}. 

Before leaving this example we note that there is no extra phase
arising from performing the Gaussian integrals in this limit. This is
completely in line with the observations of Bar-Natan and Witten
\cite{Bar-Natan-Witten}. 

\subsection{The $k \longrightarrow 0$ Limit}

The $k \longrightarrow 0$ limit greatly simplifies
the action of the finite dimensional theory that we arrive at, as one can see in
(\ref{const-action-2}).  

This limit is interesting from the gravitational view point as, if one
considers $\mathcal{B}$ to be a (possibly singular) dreibein, the action
is that of gravity with a cosmological constant \cite{WBF}. Predominantly,
the manifolds of current interest for applications to quantum gravity
are non-compact and require considerations like boundary conditions,
an issue we do not discuss here. However, there is also, via the 3d-3d
correspondence at $k=0$ \cite{Pei-Ye}, a relationship between complex
Chern-Simons theory and the superconformal index of an associated 3d
$N=2$ theory. This comparison requires one to take $s$ purely imaginary
and thus requires us to analytically continue our results to that regime.

\section{Contact 3-Manifolds and Gauge Conditions}\label{section-gf}

Technically we will need to make use of two
facts having to do with gauge fixing. The first
fact is that one can impose that the component
$\mathscr{\phi} $ of the connection $\mathscr{C}$
along the fibre direction of the Seifert manifold can be gauge fixed to be
constant along the fibre. The second fact is that one may then, at a
price, conjugate $\mathscr{\phi} $ into a Cartan sub-algebra of the
Lie algebra of the compact group \cite{btdia}. Once the gauge fixing
has been performed one finds that the evaluation can be pushed down to
a problem in two dimensional Abelian gauge theory. One decomposes the connection
in terms of Fourier modes (along the fibre) and then one integrates over all the
massive Fourier modes of the connection leaving an effective theory on
the (orbifold) base. Obviously the integration over the infinite
number of Fourier components of the connection requires its own set of
techniques.

To set notation we let $\lg$ be the Lie algebra of the compact group
$G$, $\lt$ a Cartan subalgebra and $\lk$ the complement to $\lt$ in
$\lg$,
\be
\lg = \lt \oplus \lk
\ee
Likewise the complexification of
$\lg$, $\lg_{\mathbb{C}}$, is then the Lie algebra of
$G_{\mathbb{C}}$ and we let $\lt_{\mathbb{C}}$ be the complexification of $\lt$
so that we have
\be
\lg_{\mathbb{C}} = \lt_{\mathbb{C}} \oplus \lk_{\mathbb{C}}
\ee

All 3-manifolds admit a contact structure, that is $M$ can be equipped
with a globally defined
one-form $\kappa$ such that $\kappa \wedge d\kappa \neq 0$. When $M$
is considered to 
be a contact manifold $(M,\, \kappa)$,(and $M$ now not necessarily a 
$\mathbb{Q}$HS), there is a natural decomposition
of the tangent bundle and consequently of the cotangent bundle so that
we may decompose connections as
\be
\mathscr{A} = A + \kappa \, \phi, \;\;\; \mathscr{B}= B + \kappa \,
\lambda \,.\label{decomp1}
\ee
Given a contact structure one also has a Reeb vector field $K$ such
that
\be
\iota_{K} \kappa =1, \;\;\; \iota_{K} d\kappa =0
\ee
When $M$ is Seifert 3-manifold we take $\kappa$ to be the connection
1-form and $K$ the generator of $S^{1}$ on $M$.

Given this decomposition (\ref{decomp1}) the action becomes
\bea
I(k, \, s) &=& \frac{k}{4\pi}\int_{M}\kappa\wedge\Tr{\left[  -A\wedge
  L_{\phi}A + B\wedge L_{\phi} B -2 \lambda\, d_{A}B + 2 \phi\, dA
+ d\kappa  \left( 
    \phi^{2}-  \lambda^{2}  \right)  \right]}
\nonumber \\ 
&  -& \frac{s}{2\pi}
\int_{M} \kappa \wedge \Tr{ \left[  -B\wedge L_{\phi}\, A + B \wedge
    d\phi + \lambda \wedge dA
  + \frac{1}{2}  B\wedge [\lambda, \, B] \right. } \nonumber\\
& & \left. \;\;\;\;\; \;\;\;\;\; - \frac{1}{2} A\wedge
  [\lambda, \, A ] + d\kappa \, 
    \lambda \phi\right]
\eea
where the twisted Lie derivative is
\be
L_{\phi} \equiv \iota_{K}\circ d_{\phi} + d_{\phi} \circ \iota_{K}
\ee
and the covariant derivative that enters is twisted along the flow
\be
d_{\phi} = d + \kappa \phi
\ee


Now we impose the condition that
\be
\iota_{K}\,  d \, \iota_{K} \mathscr{C}=0 \Rightarrow \iota_{K}
d \phi=0, \;\; \iota_{K}d \lambda =0 \label{constant}
\ee
Furthermore, we also impose the conditions that along the fibre the
connection is in the Cartan subalgebra of $G_{\mathbb{C}}$,
\be
\iota_{K}\mathscr{C}^{\lk_{\mathbb{C}}}=0\; \Rightarrow
\phi^{\lk}=0 , \; \;\; \lambda^{\lk}=0
\ee
We will impose these conditions in a unitary manner. Let $\mathscr{D}$
be a $\lg_{\CC}$-valued 0-form and add to the action
\be
\int_{M}\Tr{\left[\mathscr{D}*\iota_{K}d\iota_{K}\mathscr{C}
 +
 \overline{\mathscr{D}}*\iota_{K}d\iota_{K}
 \overline{\mathscr{C}}\right]} 
\ee
as Lagrange multiplier fields imposing (\ref{constant}). Notice that
this is a real choice and 
that the components of $\mathscr{D}$ that do not depend on the
fibre direction of $M$ do not enter. We let $\mathscr{E}$ and $\mathscr{F}$ be
$\lg_{\CC}$-valued  0-form ghost fields, so that the ghost
action becomes
\be
\int_{M}\Tr{\left[\mathscr{E} *\iota_{K}d\iota_{K}
    d_{\mathscr{C}} \mathscr{F} + \overline{\mathscr{E}}
    *\iota_{K}d\iota_{K} 
    d_{\overline{\mathscr{C}}} \overline{\mathscr{F}} \right]}
\ee
As long as we agree to not include $L_{K}$ zero modes we may simplify
the ghost action to,
\be
\int_{M}\Tr{\left[\mathscr{E} *\iota_{K}
    d_{\mathscr{C}} \mathscr{F} + \overline{\mathscr{E}} *\iota_{K}
    d_{\overline{\mathscr{C}}} \mathscr{F} \right]} \label{ghosts}
\ee

We also want to impose that the now constant fields lie in the Cartan
subalgebra. We use the constant $\lk_{\mathbb{C}}$ part of $\mathscr{D}$ to
do that,
\be
\int_{M}\Tr{\left[\mathscr{D}*\mathscr{C} +
    \overline{\mathscr{D}}*\overline{\mathscr{C}} \right]}
\ee
The ghost terms are then correctly incorporated in (\ref{ghosts})
where we understand that the only components of the ghosts which do
not appear are those constant along the fibre and simultaneously take values
in the Cartan subalgebra. 

\section{Seifert Rational Homology Spheres}\label{section-seifert}

The 3-manifolds of interest are circle $V$-bundles over 2 dimensional
orbifolds $\Sigma$ of genus $g$ -shortly we will fix on the base
being $\mathbb{P}^{1}$ with $N$ orbifold points. The Seifert
manifold is written as $M[\deg{\mathcal{L}}, \, g, \, (a_{1}, \,
b_{1}), \dots , (a_{N}, \, b_{N})]$ where the $a_{i}$ are the
isotropies of the orbifold points, the $b_{i}$ are the weights of the
line $V$-bundle at the orbifold points and $\deg{\mathcal{L}}$ is the
degree of that line bundle. The local model at each orbifold point,
for the associated line V-bundle, is
\be
(z, w) \simeq (\zeta.z, \zeta^{b}.w), \;\;\; \zeta^{a}=1
\ee
The Seifert manifold is smooth if
$\mathrm{gcd}(a_{i}, \, b_{i}) =1 $ for each $i$. It is an
$\mathbb{Z}$HS iff the line bundle $\mathcal{L}_{0}$ that defines it
satisfies
\be
g=0, \;\;\; c_{1}(\mathcal{L}_{0})= \pm \frac{1}{a_{1} \dots
  a_{N}} \label{zhs} 
\ee
one consequence of these conditions is that $\mathrm{gcd}(a_{i}, \,
a_{j}) =1 $ for $i\neq j$.
If one takes a tensor power of this line V-bundle,
$\mathcal{L}_{0}^{\otimes d}$ then the Seifert manifold is a
$\mathbb{Q}$HS with
\be
g=0, \;\;\; c_{1}(\mathcal{L}_{0}^{\otimes d})= \pm \frac{d}{a_{1} \dots
  a_{N}} \label{qhs} 
\ee
and
\be
|d| = |\mathrm{H}_{1}(M, \, \mathbb{Z})|
\ee

We want to make contact with the $L(p,\, q)$ notation used for Lens
spaces. It is a theorem that all Seifert manifolds satisfying (\ref{qhs}) with
$N\leq 2$ are Lens spaces (see \cite{Orlik} page 99). Indeed all Seifert
manifolds satisfying (\ref{zhs}) with $N\leq 2$ are $S^{3}$. Fix $N=2$
and recall that
for a Lens space $L(p,\, q)$ the order  of the first integral homology
group is
$|p|$ so that we may identify $p$ with $d$ in (\ref{qhs}). This gives
us a formula, namely
\be
p= a_{1}a_{2} c_{1}(\mathcal{L}_{0}^{\otimes d}) = a_{1}a_{2}\left( n +
  \frac{b_{1}}{a_{1}} + \frac{b_{2}}{a_{2}}\right)
\ee
where $n = \deg{\mathcal{L}_{0}^{\otimes d}}$. The $q$ of the Lens space is then
determined in the following way. Let $r$ and $s$ be integers
satisfying $ra_{1}-s(na_{1}+b_{1})=1$ then
\be
q = ra_{2}-sb_{2}
\ee

\section{Calculation of the Determinants}\label{section-2d}
The calculation of the ratio of the determinants in the $G_{\mathbb{C}}$ theory
is very similar to that for gauge group $G$ \cite{BT-Seifert}. We will
concentrate mostly on the differences with that exposition. 

The ratio of determinants that we wish to evaluate is, see (5.5) of
\cite{BT-Seifert}, 
\be
\frac{\Det{\left(i \widetilde{L}_{(\phi, \, \lambda)}
    \right)_{\Omega^{0}(M, \, \lk) \otimes \Omega^{0}(M, \, \lk)
    }}}{\sqrt{\Det{\left(*\kappa \wedge i\widehat{L}_{(\phi, \,
          \lambda) }\right)_{\Omega^{1}_{H}(M, \, \lk) \otimes
        \Omega^{1}_{H} (M, \, \lk)
    } } }}\label{rat-det}
\ee
where the operator $* \kappa \wedge i\widehat{L}_{(\phi, \, \lambda)}$
  acts on horizontal $\lk$ valued forms
\be
* \kappa \wedge i\widehat{L}_{(\phi, \, \lambda)}: \Omega^{1}_{H}(M, \,
  \lk) \otimes  \Omega^{1}_{H} (M, \, \lk)\longrightarrow
  \Omega^{1}_{H}(M, \, \lk) \otimes  \Omega^{1}_{H} (M, \, \lk)
\ee
while the ghost operator is 
\be
i \widetilde{L}_{(\phi, \, \lambda)}: \Omega^{0}(M, \, \lk) \otimes
\Omega^{0}(M, \, \lk) \longrightarrow \Omega^{0}(M, \, \lk) \otimes
\Omega^{0}(M, \, \lk) 
\ee
and neither of these are diagonal, though both are Hermitian. As usual
we decompose the space of forms into modes along the circle direction
as
\be
\Omega^{1}_{H} (M, \, \lk) = \bigoplus \Omega^{1} (\Sigma, \,
\mathcal{L}^{\otimes n} \otimes V_{\lk}), \;\;\; \Omega^{0} (M, \,
\lk) = \bigoplus \Omega^{0} (\Sigma, \, 
\mathcal{L}^{\otimes n} \otimes V_{\lk}),
\ee
and we also decompose the charge space into roots
\be
V_{\lk}= \oplus_{\alpha} V_{\alpha}
\ee

On expanding the fields $A$ and $B$ in terms of Fourier modes and
charges (concentrating on those that take values in the complement of the Cartan
subalgebra) we may write the operators in matrix form as
\bea
& & \left. * \kappa \wedge i\widehat{L}_{(\phi, \,
    \lambda)}\right|_{\Omega^{1}(\Sigma , \,
\mathcal{L}^{\otimes n} \otimes V_{\alpha}) \oplus \Omega^{1}(\Sigma , \,
\mathcal{L}^{\otimes n} \otimes V_{\alpha})}\nonumber \\
& & \;\;\;\;\;\;\; =  \frac{1}{\sqrt{(k^{2}+s^{2})}}
\left( \begin{array}{ll}
k(n + i \alpha(\phi))- si \alpha(\lambda) & -s(n+i\alpha(\phi))- ki
\alpha(\lambda) \\
-s(n+i\alpha(\phi))- ki\alpha(\lambda) & -k(n+i\alpha(\phi))+
si\alpha(\lambda) 
\end{array} \right) \label{1-form-operator}
\eea
the factor of $(k^{2}+s^{2})^{-1/2}$ is there so that the measure for
the fields is simply $\prod_{n}dA_{n}\, dB_{n}$, however, there is
still a factor of $\sqrt{t\overline{t}}$ for each of the modes of the
fields which we have not
integrated. To put the ghost
operator in a similar form we expand as before
\be
\mathscr{E} = E+i \mathbb{E}, \;\;\; \mathscr{F} = F + i\mathbb{F}
\ee 
and the ghost operator is
\be
\left. i \widetilde{L}_{(\phi, \, \lambda)}\right|_{\Omega^{0}(\Sigma , \,
\mathcal{L}^{\otimes n} \otimes V_{\alpha}) \oplus \Omega^{0}(\Sigma , \,
\mathcal{L}^{\otimes n} \otimes V_{\alpha})}
 = 
\left( \begin{array}{cc}
n+i \alpha(\phi) & i\alpha(\lambda)\\
i\alpha(\lambda) & - n- i\alpha(\phi)
\end{array} \right) \label{o-form-operator}
\eea

Given an operator $T$ one may define the absolute value of its
determinant following \cite{Schwarz} as the positive root
$\sqrt{\det{\left(TT^{\dagger}\right) }}$. The operators in question
are Hermitian so we want to take their square. The matrix square of
(\ref{1-form-operator}) is
\be
\left[(n+ i\alpha(\phi))^{2}+ (i\alpha(\lambda))^{2}  \right] \, .\,
I_{2\times 2}
\ee
where $I_{2\times 2}$ is the two by two identity matrix and the matrix
square of (\ref{o-form-operator}) is  
\be
\left[ (n+ i\alpha(\phi))^{2}+ (i\alpha(\lambda))^{2}  \right]\, .\,
I_{2 \times 2}
\ee
Though these last two expressions appear to be the same one must recall
that they act on different spaces of forms. Nevertheless, they may be
`holomorphically factorized' as
\be
\left[ n+ i\alpha(\phi)+ \alpha(\lambda) \right]\,
.\,\left[ n+ i\alpha(\phi)-\alpha(\lambda) \right]\,
.\,I_{2 \times 2} 
\ee

Following through the same reasoning as presented in section 5 of
\cite{BT-Seifert} we find that the absolute value of the ratio of
determinants (\ref{rat-det}) is simply
\be
\sqrt{\tau_{M}(\phi+i\lambda,\, a_{1}, \dots , a_{N})}\, . \,
 \sqrt{\tau_{M}(\phi-i\lambda,\, a_{1}, \dots , a_{N}) }
\ee
where
\be
\tau_{M}(\Phi,\, a_{1}, \dots , a_{N}) =
\tau_{S^{1}}(\Phi)^{2-2g-N}\, . \, \prod_{i=1}^{N}
\tau_{S^{1}}(\Phi/a_{i}) \label{RST-decomp}
\ee
and the Ray-Singer Torsion of the circle being
\bea
\tau_{S^{1}}(\Phi) &=& \prod_{m, \, \alpha}\left(2\pi m +i
  \alpha(\Phi) \right) \nonumber\\
&=& \prod_{\alpha} 2\sin{\left(i\alpha(\Phi) \right)} \label{RST-circle}
\eea

Incidentally, this calculation implies that the Ray-Singer torsion for
non-unitary flat connections, $\widehat{\tau}_{M}$, factorises
holomorphically, at least for 
these Seifert manifolds and for these particular Abelian connections.

We have not explicitly calculated the phase of the determinants but
one way to see that there is no overall phase correction is to note
that both matrices (\ref{1-form-operator}) and (\ref{o-form-operator})
are traceless. Being traceless means that the two eigenvalues are of
the same absolute value but with the opposite sign and so cancel each
others contribution in the evaluation of the $\eta$ invariant.

Having integrated out all the non-zero modes in the $S^{1}$ direction of $M$
and all those zero modes in the $\lk_{\mathbb{C}}$ part of the Lie algebra, we
are left with an Abelian theory on the orbifold base of the fibration,
\be
Z_{M}[t, \, \overline{t}] = \sum_{\mathbf{n}} \int D\Phi\,
D\overline{\Phi} \, D\mathbb{A}\, D\overline{\mathbb{A}} \;
\exp{\left(iI_{\Sigma}^{\mathbf{n}}(t, \, \overline{t}) \right)}\;\;
\widehat{\tau}_{M}^{1/2}(\Phi;\, a_{1}, \dots , a_{N})\label{ab-partition}
\ee
Here we have set $\Phi = \phi + i \lambda$, $\mathbb{A} = A + iB$, 
\be
\widehat{\tau}_{M}(\Phi;\, a_{1}, \dots , a_{N})= \tau_{M}(\Phi,\,
a_{1}, \dots , a_{N}) \, . \, \tau_{M}(\overline{\Phi},\, a_{1}, \dots
, a_{N}) \label{RST-complex}
\ee
and the action is
\bea
I_{\Sigma}^{\mathbf{n}}(t, \, \overline{t}) & = & 
\frac{t}{4\pi}\int_{\Sigma}\Tr{\left(\Phi \wedge
    (F_{\mathbb{A}}+ i2\pi\mathbf{n}\,  \omega)\right)} +
\frac{\overline{t}}{4\pi}\int_{\Sigma}\Tr{\left( \overline{\Phi} \wedge
    (F_{\overline{\mathbb{A}}}-i2\pi \mathbf{n}\,  \omega)\right)}\nonumber
\\
& & \;\;\;\; + \frac{t}{8\pi}\int_{\Sigma} \Tr{\left( \Phi^{2}\right)} \wedge \omega +
\frac{\overline{t} }{8\pi}\int_{\Sigma} \Tr{\left(
    \overline{\Phi}^{2}\right)}\wedge \omega \label{ab-action}
\eea
The sum over $\mathbf{rank}(G)$ $U(1)$ bundles is there to take into
account the non-triviality of the Abelian gauge fixing, so that $A$ is
a connection on a product of trivial line bundles over $\Sigma$. We
have not specified the range of the summation as that depends on the
choice of 3-manifold. In the next section we will pick a certain class
of 3-manifolds.

Notice that
the form of (\ref{ab-partition}) with action (\ref{ab-action}) still
suggests a `holomorphic' factorization of the path integral,
\be
Z_{M}[t, \, \overline{t}] \stackrel{?}{=} \sum_{\mathbf{n}}\,
Z_{\Sigma}^{\mathbf{n}}[t]\, . \,
\overline{Z_{\Sigma}^{\mathbf{n}}[t]} \label{first-factor}
\ee
with
\be
Z_{\Sigma}^{\mathbf{n}}[t]= \int D\Phi\,D\mathbb{A}\,
\exp{\left(
iI_{\Sigma}^{\mathbf{n}}(t) \right)}\;\;\tau_{M}^{1/2}(\Phi,\,
a_{1}, \dots , a_{N}) ,
\ee
\be
I_{\Sigma}^{\mathbf{n}}(t) = \frac{t}{4\pi}\int_{\Sigma}\Tr{\left(\Phi \wedge
    (F_{\mathbb{A}}+ i2\pi\mathbf{n}\,  \omega)\right)}
+\frac{t}{8\pi}\int_{\Sigma} \Tr{\left( \Phi^{2}\right)} \wedge \omega  
\ee
We could have included a phase factor that only depends on the gauge
group and on the 3-manifold $M$ in the definition of the holomorphic
partition function -providing it cancels against a similar phase from
the anti-holomorphic partition function.

\section{Finite Dimensional Integrals}\label{section-finite}

At this point we must make an assumption about the 3-manifold $M$. The
Abelian theory (\ref{ab-partition}) has zero modes in that
elements $B \in \mathrm{H}^{1}(\Sigma, \, i\lt) $ do
not appear in the argument of the path integral. There are also $A$
zero modes but large gauge transformations would ensure that they are
compact directions. To avoid this issue we assume that the genus of
$\Sigma$ is zero.

With this assumption the path integral (\ref{ab-partition}) can be
simplified further on noting that integration over $\mathbb{A}$ and
$\overline{\mathbb{A}}$ imply that $d\Phi=0$. 

\be
Z_{M}[t, \, \overline{t}] \simeq \sum_{\mathbf{n}} \int_{\lt_{\mathbb{C}}} 
\exp{\left(iI^{\mathbf{n}}(t, \, \overline{t}) \right)}\;\;
\widehat{\tau}_{M}^{1/2}(\Phi;\, a_{1}, \dots , a_{N})\label{finite-partition}
\ee
with
\be
I^{\mathbf{n}}(t, \, \overline{t}) =   \Tr{\left(
i\frac{t}{2}\Phi \mathbf{n}  -i
\frac{\overline{t}}{2}\overline{\Phi} 
    \mathbf{n}\right)} -c_{1}(\mathcal{L})\Tr{\left(
\frac{t}{8\pi} \Phi^{2} +
\frac{\overline{t} }{8\pi}
    \overline{\Phi}^{2} \right)}\label{const-action}
\ee
As $c_{1}(\mathcal{L})\in \mathbb{Q}$ we set
$c_{1}(\mathcal{L})=d/P$. The finite dimensional `action'
(\ref{const-action}) transforms as 
\be
I^{\mathbf{n}}(t, \, \overline{t}) \longrightarrow I^{\mathbf{n}}(t,
\, \overline{t}) - 2\pi k P\,\Tr{\left(\mathbf{n} \, \mathbf{r}
  \right)} -  \pi k P d\,\Tr{\left( \mathbf{r} \, \mathbf{r}\right) }
\ee
under
\be
\Phi \longrightarrow \Phi + 2\pi  i \mathbf{r} P , \;\;\; \mathbf{n}
\longrightarrow \mathbf{n} + d \mathbf{r}\label{shift-sym}
\ee
and provided that $\Tr{\left( \mathbf{r} \, \mathbf{r}\right) }\in 2\mathbb{Z}$
the exponential in (\ref{finite-partition}) is invariant under these
transformations. On taking $P= \prod a_{i}$ the Ray-Singer torsion
$\widehat{\tau}_{M}$ is also invariant under (\ref{shift-sym}). These
transformations correspond to shifts by the integral lattice $I$ of
$\lt$ (not the complexified integral lattice for which there
is no such invariance). 

One should also consider the action of the Weyl group which acts
naturally on $\lt$ as well as on $\lt_{\mathbb{C}}$. This is a
symmetry of finite dimensional theory as it is part of the (ungauged)
gauge group. On quotienting out by these residual
invariances we have either of two formulae. One by using the symmetry
to reduce the integrals to $\lt_{\mathbb{C}}/I\rtimes W$, the other to
restrict the range of the summation. Consequently,
\be
Z_{M}[t, \, \overline{t}] =
\sum_{\mathbf{n}}
\int_{\lt_{\mathbb{C}}/I\rtimes W} 
\exp{\left(iI^{\mathbf{n}}(t, \, \overline{t}) \right)}\;\;
\widehat{\tau}_{M}^{1/2}(\Phi;\, a_{1}, \dots , a_{N})\label{finite-partition2}
\ee
or
\be
Z_{M}[t, \, \overline{t}] = \frac{1}{|W|}\, .\, 
\sum_{1
  \leq\mathbf{n}\leq \mathbf{d}}
\int_{\lt_{\mathbb{C}}} 
\exp{\left(iI^{\mathbf{n}}(t, \, \overline{t}) \right)}\;\;
\widehat{\tau}_{M}^{1/2}(\Phi;\, a_{1}, \dots , a_{N})\label{finite-partition3}
\ee

Clearly (\ref{finite-partition}), (\ref{finite-partition2}) and
(\ref{finite-partition3}) appear
to have the holomorphic decomposition
\be
Z_{M}[t, \, \overline{t}] \stackrel{?}{=}
\frac{\exp{\left(-i\pi/2\right)}}{|W|} \sum_{\mathbf{n}}\,
Z_{\lt_{\Gamma}}^{\mathbf{n}}[t]\, . \,
\overline{Z_{\lt_{\Gamma}}^{\mathbf{n}}[t]} 
\ee
with
\be
Z_{\lt_{\Gamma}}^{\mathbf{n}}[t]= 
\int_{\lt_{\Gamma}} 
\exp{\left(  i I^{\mathbf{n}}_{\lt_{\mathbb{C}}}[t]\right)}\,
\tau_{M}^{1/2}(\Phi;\, a_{1}, \dots , a_{N}) \label{half-partition}
\ee
where 
\be
I^{\mathbf{n}}_{\lt_{\mathbb{C}}}[t]= \Tr{\left(
i\frac{t}{2}\Phi \mathbf{n}   -c_{1}(\mathcal{L})
\frac{t}{8\pi} \Phi^{2}  \right)}
\ee
and there is either a restriction on the range of summation or on the
integration. We could have also introduced a phase into the
definition, but prefer not to. This has consequences for the
interpretation of the factorization formula as we will see. The
contour $\lt_{\Gamma}$ needs to be defined. In the case of Lens spaces
one may take it to be $\lt_{\Gamma}=\lt$.

Once one has factorized it is no longer necessarily true that
\be
Z_{\lt_{\Gamma}}^{\mathbf{n}+ d\mathbf{r}}[t]=Z_{\lt_{\Gamma}}^{\mathbf{n}}[t]
\ee
as the symmetry (\ref{shift-sym}) holds through a cancellation between
the holomorphic and anti-holomorphic parts of the action, so that the
best one can hope for is
\be
Z_{\lt_{\Gamma}}^{\mathbf{n}+ d\mathbf{r}}[t]=\pm
Z_{\lt_{\Gamma}}^{\mathbf{n}}[t] 
\ee


The Ray-Singer Torsion $\tau_{M}(\Phi)$ (\ref{RST-decomp}) diverges at
the zeros of the sine function (\ref{RST-circle}) when the number of
orbifold points is such that $N + 2g >2$. This means that the torsion
$\widehat{\tau}_{M}$ will also have poles. The zeros of $
\sin{\left(x+iy\right)}$ are at $(x, \, y)= (m \pi,\, 0)$ for $m \in
\mathbb{Z}$. 

If one first makes use of the symmetry
(\ref{shift-sym}) to restrict the range of integration of $\Phi$,
actually of $\phi$, then the integrals to be performed over $\phi$
range from over $-\pi P \leq \phi \leq \pi P$ for each component in
the Cartan subalgebra. The sum over the integers $\mathbf{n}$ then
implies that $k\phi - s \lambda = \mathbf{r} \pi$ for $\mathbf{r} \in
\mathbb{Z}^{\rk}$. The singularities occur at $(\phi, \,
\lambda)=(\mathbf{m}\pi , \, 0)$ whence at
$\phi = \mathbf{r} \pi/k = \mathbf{m}\pi$ so that the divergences are
at $k|\mathbf{r}$. For compact gauge group $G$ the divergences are at
$(k+ c_{\lg})|\mathbf{r}$ (but we do not see the shift in the level in
$G_{\mathbb{C}}$). We may give the charged gauge fields (those in the
orthocomplement of $\lt_{\mathbb{C}}$) a small mass while preserving
the maximal torus symmetry of the action. As for Chern-Simons theory with
compact gauge group $G$ this means that at $\lambda=0$ we have the
ratio
\be
\sin^{2}{\left(i\alpha(\phi)\right)}/\sin^{N}{\left(i\alpha(\phi)+
    \eps\right)}
\ee
which is non-singular on the walls, indeed vanishes there.

\section{Summing over Flat Connections}\label{section-flat}

A comparison of (\ref{half-partition}) with our formula for compact
$G$ (6.2) in  \cite{BT-Seifert} shows that we have 
\be
Z_{G}[M, k] = \sum_{\mathbf{r}}\frac{ \exp{\left(4\pi i \Phi(\mathcal{L})
  \right)}}{|W|}\, .\, Z^{\mathbf{r}}_{\lt}[M, 2(k+ 
 c_{\lg})] \label{real-version}
\ee
while the formula for the partition function of Chern-Simons theory
with gauge group $G_{\mathbb{C}}$ is
\be
Z_{G_{\mathbb{C}}}[M, k] = \frac{\exp{\left(-i\pi/2\right)}}{|W|}
\sum_{\mathbf{r}}  
Z^{\mathbf{r}}_{\lt_{\Gamma}}[M, t ]\, . \,
Z^{\mathbf{r}}_{\lt_{\Gamma}}[M, \widehat{t} ] 
\ee
In passing to (\ref{real-version}) one is losing some information as
$t$ is a Gauss integer, so it has integral real part, while $2(k+ 
 c_{\lg})$ is even.

The summation in the factorization formulae is over an integer which,
presently, appears to have no geometric significance from the point of
view of the original theory. Here we will show that, quite
miraculously, the summation is at least in a limit over certain flat
connections. To establish
this relationship we will need to take a closer look at the large $s$ limit.

In components the action of the finite dimensional integral for any
Seifert $\mathbb{Q}$HS $M$ is 
\be
I^{\mathbf{n}}(t, \, \overline{t}) =   \Tr{\left[
(k\phi -s\lambda)i\mathbf{n} \right]} + c_{1} (\mathcal{L}) \Tr{
\left[\frac{k}{4\pi} 
 (\phi^{2} -\lambda^{2}) - \frac{s}{2\pi}\phi \lambda
\right]}\label{const-action-2} 
\ee
In order to compare with the large $s$ limit of Section
\ref{various-limits} we must scale $\lambda \longrightarrow \lambda/s$
in which case the action goes over to
\be
I^{\mathbf{n}}(t, \, \overline{t}) \longrightarrow   \Tr{\left[
(k\phi -\lambda)i\mathbf{n} \right]} + c_{1} (\mathcal{L}) \Tr{
\left[\frac{k}{4\pi} 
 \phi^{2} - \frac{1}{2\pi}\phi \lambda
\right]}\label{const-action-3} 
\ee
Unfortunately, the scaling multiplies the integral with a prefactor of
$|s|^{-\mathbf{rk}(G)}$, which in the limit means that the path
integral vanishes. In order to pass to the $BF$ theory we must
multiply the path integral by $|c_{1}(\mathcal{L}_{0}^{\otimes
  d})s|^{\mathbf{rk}(G)}$ and we presume 
that this has been done.

This scaling also turns the argument of the Ray-Singer Torsion from $\Phi$
to $\phi$, 
\be
\widehat{\tau}^{1/2}_{M}(\Phi) \longrightarrow
  \tau_{M}(\phi) 
\ee

As $\lambda$ now only appears in the action
(\ref{const-action-3}), the integral over $\lambda$ imposes the
condition
\be
c_{1}(\mathcal{L}_{0}^{\otimes d})\, \phi= -2\pi i \mathbf{n}\nonumber
\ee
that is
\be
\phi= 2\pi i \mathbf{n} \frac{P}{d}
\ee
as a delta function constraint. In this way we obtain for the
partition function
\be
Z_{M}[t, \, \overline{t}] \stackrel{s\rightarrow
  \infty}{\longrightarrow} 
\sum_{\mathbf{n}}  \,
\exp{\left(i\pi k P\Tr{\mathbf{n}^{2}}/d)\right) }\, .\,
\tau_{M}(2\pi i P\mathbf{n}/d) \label{large-s-M}
\ee

Some things about this result 
requires
comment. Note that we do not 
attempt to sum over flat connections at all, yet the final sum may be
interpreted as that over flat Abelian connections. We have introduced
a sum over non-trivial $U(1)$ bundles in (\ref{ab-action}) which may
be viewed as arising from background connections with
values in the Cartan subalgebra of $G$ which are Yang-Mills
connections (that is they satisfy the Yang-Mills equations) with the
explicit connection being
\be
A = 2\pi P\frac{\mathbf{n}}{d} \kappa, \;\;\; F_{A}= 2\pi \mathbf{n}
\omega\;\; \mathrm{and}\;\; d_{A}*F_{A}=0\label{YM-connection}
\ee
As explained by Atiyah and Bott \cite{Atiyah-Bott} section 6 and, in a
version closer to our needs, in section 5 of Beasley and Witten
\cite{Beasley-Witten} there is a close relationship between the space
of flat connections on $M$ a circle bundle over $\Sigma$ and
Yang-Mills connections on $\Sigma$. To explain this we need the
generators and relations defining the various fundamental groups. As
$M$ is a $\mathbb{Q}$HS the generators of $\pi_{1}(M)$, $c_{i}$, $i 
= 1, \dots , N$ and $h$ satisfy
\be
[c_{i}, \, h]=1, \;\;\; c_{j}^{a_{j}}h^{b_{j}} =1, \;\;
\mathrm{and}\;\; \prod_{j=1}^{N}c_{j}= h^{n} \label{relations}
\ee
clearly $h$ is central and when $h=1$ these relations reproduce those
of the fundamental group of the orbifold $\Sigma$. Yang-Mills
connections on $\Sigma$ 
correspond to having a central extension of the fundamental group
according to \cite{Atiyah-Bott}, while in \cite{Beasley-Witten} it is
noted that as
\be
1\longrightarrow \mathbb{Z} \longrightarrow \pi_{1}(M) \longrightarrow
\pi_{1}(\Sigma) \longrightarrow 1 
\ee
that the extra generator $h$ in $\pi_{1}(M)$ then provides that
extension. In turn the Yang-Mills connection (\ref{YM-connection}) provides us
with a representation of the extension $h$
\be
\rho(h) = \exp{\left( 2\pi \oint_{S^{1}}(P\mathbf{n}/d)\,
    \,  \kappa\right) } = \exp{\left(2\pi P \mathbf{n}/d\right)}\label{our-rep}
\ee
Given an $m$-dimensional representation,
\be
\rho: \pi_{1}\left(M(n,0, (a_{1},b_{1}), \dots (a_{N},b_{N})) \right)
\longrightarrow \mathrm{GL}_{m}(\mathbb{F})
\ee
the Reidemeister torsion for a Seifert manifold $M$ is given by
\cite{Kitayama} (see Lemma 4.3 there and note our definition of the
torsion is the
inverse of the one used there)
\be
\tau_{M
  }\left(\rho\right) = \det{\left( \rho(h)-I\right)}^{2-N}
  \prod_{i=1}^{N} \det{\left( \rho(h^{s_{i}}c_{i}^{r_{i}})-I\right)}
\ee
where 
\be
a_{i}s_{i}-b_{i}r_{i}=1\label{asbr}
\ee
Our representation (\ref{our-rep}) is into the diagonal matrices so
that by (\ref{asbr}) and the
homotopy relations (\ref{relations}) we have that 
\be
\rho(h^{s_{i}}c_{i}^{r_{i}}) = \rho(h)^{s_{i}-r_{i}b_{i}/a_{i}}=
\rho(h)^{1/a_{i}} 
\ee
so that
\be
\tau_{M
  }\left(\rho\right) =
  \prod_{\alpha}4\sin{\left(2\pi P\alpha(\mathbf{r})/d\right)}^{2-N} 
  \prod_{i=1}^{N} \sin{\left(2\pi P\alpha(\mathbf{r})/da_{i} \right)}
\ee
We are, therefore, able to understand the partition function
(\ref{large-s-M}) as being given by a sum over flat connections as
required by (\ref{large-s}), however, not a sum over all possible flat
connections but rather over a class of Abelian flat connections. To
complete the comparison we also note that the exponent in
(\ref{large-s-M}) is just the evaluation of the Chern-Simons action on
the flat connection of interest.

In summary one interprets the summation as being over Abelian connections
for all allowed $M$ then (\ref{Witten-factor}) holds where the representations
$\rho$ arise from very special flat connections, namely those that are
Abelian, are in the Cartan subalgebra of $G$ and correspond to the
generator $h$. For such $\rho$ we would then also have that $n_{(\rho,
\overline{\rho})}=1$, and that the factors $Z_{M}^{\rho}[t]$ are, in
some way, the complex Chern-Simons theory exactly evaluated about our
preferred Abelian connection. In the following section we show in
which sense that this is the case for Lens spaces.

\section{Some Calculations on $L(p,q)$
with $G_{\mathbb{C}}=SL(2,
  \mathbb{C}) $}\label{section-calcs}

In this section we will derive some exact formulae for the
partition function on $L(p,q)$ with complex gauge group
$SL(2,\mathbb{C})$. A bit further on we consider the expectation value
of some knots too. These calculations allow one to see the explicit
dependence on framing. We represent these Lens spaces as Seifert
manifolds either with one or two exceptional fibres.

\subsection{Holomorphic Factorization}

According to (\ref{Gaussian}) and
(\ref{contours}) the partition function factorises with holomorphic part
\be
Z_{\lt_{\Gamma}}^{r}[L(p,q), \, t]
=
\frac{1}{\sqrt{4 \pi^{2} a_{1}a_{2}}}\int_{\Gamma}\,
\exp{\left(-itzr+i\frac{tp}{4\pi a_{1}a_{2}}z^{2}\right)} \, 
4 \sin{(z/a_{1})}\, .\, \sin{(z/a_{2})}\label{sl2c-hol}
\ee
On $S^{3}$ we have that $p=1$ which means that $r=0$ but we consider
$S^{3}$ as the Seifert 3-manifold, 
$M(n, 0, (a_{1}, b_{1}), (a_{2}, b_{2}))$ with $n$ chosen so that $c_{1}=
\pm 1/a_{1}a_{2}$. The standard Hopf fibration corresponds to
$a_{1}=a_{2}=1$, $b_{1}=b_{2}=0$ and $n=\pm 1$. As $r=0$ we
have a very simple integral to 
perform indeed. By writing the product of the sine functions as sums
of exponentials we arrive at
\be
Z_{\lt_{\Gamma}}[S^{3}, \, t]= \exp{\left(i \frac{\pi}{4} -i\frac{ \pi}{t}
    \frac{a_{1}^{2}+a_{2}^{2}}{a_{1}a_{2}}\, \right) } \,
. \, 
\frac{4}{\sqrt{t}}\, \sin{\left(\frac{2\pi}{t} \right)} 
\ee
The phase is a consequence of our choice of framing. Combining with
the anti-holomorphic part we obtain
\be
Z_{SL(2,\mathbb{C})}[S^{3},  t , \widehat{t}] = \exp{\left( -i \pi
    \frac{a_{1}^{2}+a_{2}^{2}}{a_{1}a_{2}}\left(\frac{1}{t}+
      \frac{1}{\widehat{t}} \right)\right) } \, . \, 
\frac{8}{\sqrt{t\widehat{t}}}\, \sin{\left(\frac{2\pi}{t} \right)}\,
\sin{\left(\frac{2\pi}{ \widehat{t}} \right)} 
\ee
The phase prefactor is neatly written as
\be
\exp{\left(2\pi i m (c_{L}-c_{R})/24\right)}, \;\; m =
(a_{1}^{2}+a_{2}^{2})/a_{1}a_{2} 
\ee

We now view the general Lens spaces as the Seifert Manifolds
$L(p,q)\equiv M(n,0, (a,b) )$ where $q=a$ and $p= na + b$
(\cite{Orlik} page 99). That is we let $(a_{1}, b_{1})=(1,0)$ and
$(a_{2}, b_{2}) =(q, b)$ and use (\ref{sl2c-hol}) to calculate 
\be
Z_{\lt_{\Gamma}}^{r}[L(p,q), \, t]
= \frac{2i}{\sqrt{tp}} \ex{\left(\frac{i\pi}{4}-\frac{i \pi}{tp}(q+1/q)\right) }\ex{\left(-\frac{i
      \pi}{p}tqr^{2} \right) }\sum_{\eps = \pm 1} \eps \cos{\left(\frac{2\pi
  r(q+ \eps)}{p}\right)}
\ex{\left(-\eps \frac{2\pi i}{tp} 
  \right)} \nonumber
\ee
Notice that this expression is not necessarily invariant under $r
\rightarrow r+p$. Apart from a phase and an overall factor related to
the order of the Weyl group this expression does, however, compare
favourably with that of Jeffrey \cite{Jeffrey} Theorem 3.4. on taking
$t=2(k+2)$.


As we have the explicit $s$ dependence of the path integral in terms
of known functions we may extend the formulae to the
complex $s$ plane.

\subsection{Inclusion of Knots}

There is great advantage in having different fibrations represent the same
topological space. Given a Seifert fibred 3-manifold the fibre over a
regular point of the base is a Torus knot $\mathcal{K}_{a_{1}, \,
  a_{2}}$. The type of Torus knot 
depends on the choice of fibration. This means that Wilson loops in
the vertical direction can be invariants for different knots in the
same manifold by changing the fibration. This approach was used by
Beasley in \cite{Beasley} in a study of Chern-Simons theory with knot
invariants within the context of non-Abelian localisation.
\begin{figure}[h]
\centering
\includegraphics[height=3cm]{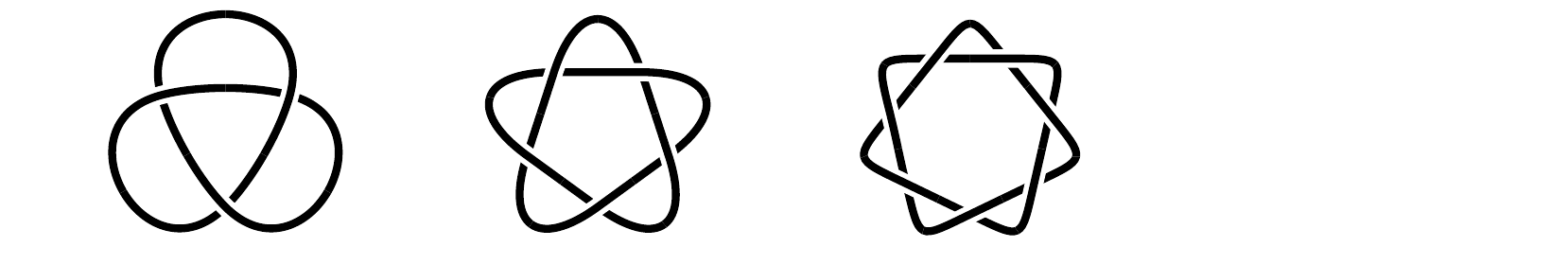}
\caption{Torus Knots which are the regular fibres of the 3-sphere
  viewed as different Seifert fibrations. These are $\mathcal{K}_{3,
    \, 2}$, $\mathcal{K}_{5, \, 2}$ and $\mathcal{K}_{7, \, 2}$
  respectively. They are the $3_{1}$, $5_{1} $ and $7_{1} $ knots in
  Rolfsen's list \cite{Rolfsen}. }
\end{figure}
In a finite dimensional
representation $R$, the possible Wilson loops along the fibre are
\be
\Tr_{R}{\left( P \, \exp{i \oint \Phi}\right)}, \;\;\;\mathrm{and}
\;\;\;  \Tr_{R}{\left(
    P \, \exp{i \oint \overline{\Phi} }\right)}
\ee
Once we have diagonalised then path ordering is no longer required
and, with the Cartan subalgebra being made up of diagonal matrices,
such traces are just finite sums of exponentials of $\Phi$ or
$\overline{\Phi}$. For example if we fix on $G_{\mathbb{C}}= SL(2,
\mathbb{C})$ then in a, finite dimensional
representation $R$ of dimension $n+1$
\be
\Tr_{R}{\left( P \, \exp{i \oint \Phi}\right)} = \frac{\sin{\left(
      (n+1)i\Phi 
    \right)}}{\sin{\left(i\Phi \right)} } = \sum_{j=-n}^{n}
\exp{\left(  j \Phi \right)}
\ee
($j/2$ is the spin). Being sums of exponentials we can use the
factorization formulae to evaluate expectation values of products of
Wilson loops, at least in the case of Lens spaces.

As before, for simplicity, we fix on $G_{\mathbb{C}}= SL(2,
\mathbb{C})$ and on $S^{3}$. We consider Torus knots of the form
$\mathcal{K}_{a_{1}, \, a_{2}}$  and to do so we
consider $S^{3}$ to be given by the Seifert fibration $M(n, 0, (a_{1},b_{1}),
(a_{2}, b_{2}))$ while ensuring that the first Chern class of the
fibration satisfies $c_{1}=\pm 1/a_{1}a_{2}$. The
$\mathcal{K}_{a_{1}, \, a_{2}}$ Torus knot is then a generic fibre
of $M(n,0, (a_{1},b_{1}),(a_{2}, b_{2}) )$. 

The factorization (\ref{Gaussian},\, \ref{contours}) extends to sums
of exponentials, 
\be
\int d^{2}z \, \exp{\left(az^{2} + b\overline{z}^{2}\right)}\, f(z)\,
g(\overline{z})  =\exp{\left(-i\pi/2\right)}\int_{\Gamma}dz \,
\exp{\left(az^{2}\right)}\, f(z)\, .\, 
\int_{\Gamma'}d\overline{z} 
\, \exp{\left(b\overline{z}^{2}\right)}\,
g(\overline{z})\nonumber 
\ee
where $f(z) = \sum_{i} \eps_{i} \exp{\left( c_{i} z\right)}$ and
$g(\overline{z}) = \sum_{j} \eta_{j} \exp{\left(d_{j}
    \overline{z}\right)}$ and
\be
\int d^{2}z \, \exp{\left(az^{2} + b\overline{z}^{2}\right)}\, f(z)\,
g(\overline{z}) = \frac{\pi}{\sqrt{-ab}}\,\sum_{i\, j}
\eps_{i}\eta_{j} \exp{\left(-
    \frac{\sum_{i} c_{i}^{2}}{4a} - \frac{\sum_{j} d_{j}^{2}}{4b}
  \right)}\label{Gaussian-sum} 
\ee

We want to use holomorphic factorization once more, so we evaluate
the non-normalised `holomorphic' expectation value of a Torus knot
$\mathcal{K}_{a_{1}, a_{2}}$,
\bea
Z_{\lt_{\mathbb{C}}}[S^{3},\, \mathcal{K}_{a_{1}, a_{2}} , \, t] &
\equiv & \sum_{j=-n}^{n}\frac{1}{\sqrt{4\pi^{2} a_{1}a_{2}}}\int_{\Gamma}\,
\exp{\left(ijz+i\frac{t}{4\pi a_{1}a_{2}}z^{2}\right)} \, 
4 \sin{(z/a_{1})}\, .\, \sin{(z/a_{2})}\nonumber\\
& = & \frac{2i}{\sqrt{t}}\sum_{\eps = \pm 1}\sum_{j=-n}^{n} \eps\,
\exp{\left(i \frac{\pi}{4} -\frac{i\pi}{a_{1}a_{2}t} (a_{1}a_{2}j + a_{1} + \eps
    a_{2})^{2}  \right)}
\eea
One can compare this with the formula for $G=SU(2)$, \cite{Morton,
  Rosso-Jones}. 

The perturbative holomorphic partition function with the inclusion of
the figure eight knot, shown in Figure 2, for an infinite dimensional
representation, was determined in
\cite{Dimofte-Gukov} but our methods, unfortunately, do not extend to
give a non-perturbative evaluation in this case.

\begin{figure}[h]\label{fig-8}
\centering
\includegraphics[height=3cm]{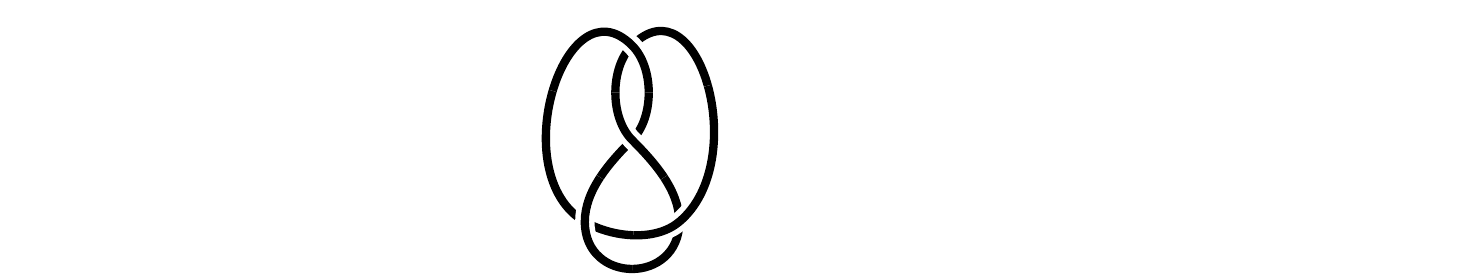}
\caption{The figure 8 knot is the simplest knot which is not a
  Torus knot. It is designated the $4_{1}$ knot in
  Rolfsen's list \cite{Rolfsen}. }
\end{figure}

\subsection{Beyond Lens Spaces}

We begin this section with an observation on the partition function in
the large $s$ limit (\ref{large-s-M}), namely that it can be written
in factorized form since we can take the root of the Ray-Singer
Torsion. This suggests, that in this limit at least, the partition
function for general Seifert manifolds also factorises.

The large $s$ limit makes no sense in (\ref{real-version}) as $s$ has
been set to zero there. What real theory does the factorization
correspond to then? We need a theory that gives us as its
`holomorphic' part the square root of the Ray-Singer Torsion evaluated at a
flat connection. One possibility is an action which includes a
Chern-Simons term and a $BF$ term to land on flat connections but such
a theory leads to the Ray-Singer Torsion and not its square root. This
situation can be remedied with the addition of
\be
\int_{M} \Tr{\left(\overline{\psi}d_{A}\psi + \rho d_{A} \rho \right) }
\ee
to the action where $\psi$ and $\overline{\psi}$ are Grassmann odd Lie algebra
valued 1-forms and $\rho$ is a Grassmann even Lie algebra
valued 1-form. This term is, by itself, not gauge invariant under gauge
transformations for the new fields, e.g. under $\rho \rightarrow \rho
+ d_{A} \sigma$ but can be compensated for by transforming $B$ (by
shifting it by a
multiple of $[\rho, \sigma]$ in this case).

We should point out that the derivation of the path integral is also
correct when the 3-manifold is $S^{2}\times S^{1}$ even though this is
not a $\mathbb{Q}$HS. The components $\phi$ and $\lambda$ are now
those along the $S^{1}$ direction and one sets $p=0$ in
(\ref{finite-partition}) and subsequent formulae. This path integral
is usually normalised to unity (as in the compact case one is counting
conformal blocks). 

The restriction that the Seifert manifold be a $\mathbb{Q}$HS comes
from the fact that we have no control over Abelian $B$ modes that are
simultaneously constant on the fibre and harmonic on the base. These
modes are not damped in the path integral and so lead to a divergence. The
corresponding $A$ modes are compact and so do not give rise to any
difficulty.

\section*{Acknowledgements}
G.T. would like to thank S. Gukov for suggesting that it might be of interest
to apply Abelianization to complex Chern Simons theory (at the Recent
Advances in Topological Quantum Field Theory meeting in Lisbon in
2012) and for many useful comments on a preliminary version of this paper. 
The work of MB is partially supported through the NCCR SwissMAP (The Mathematics
of Physics) of the Swiss Science Foundation. The
knots in Figures 1 and 2 were drawn using the PSTricks knot 
macro pst-knot created by Herbert Vo\ss.

\rnc{\Large}{\normalsize}

\end{document}